\documentstyle[twocolumn,psfig,aps]{revtex}
\begin{document}
\draft
\twocolumn[\hsize\textwidth\columnwidth\hsize\csname @twocolumnfalse\endcsname

\title{ Hole-Density
Evolution of the One-Particle Spectral Function in Doped  Ladders
}

\author{George B. Martins, Claudio Gazza, and Elbio Dagotto}

\address{National High Magnetic Field Lab and Department of Physics,
Florida State University, Tallahassee, FL 32306}

\date{\today}
\maketitle

\begin{abstract}
The spectral function $A({\bf q}, \omega)$ of doped $t-J$ ladders is 
presented on clusters with up to $2 \times 20$ sites at zero temperature
applying a recently developed technique that uses up to $\sim 6 \times 
10^6$ rung-basis states.
Similarities with photoemission results for the 2D cuprates are
observed, such as the existence of a gap at $(\pi,0)$ 
near half-filling (caused by hole pair formation) 
and  flat bands in its vicinity. These features should be observable in
ARPES experiments on ladders. 
The main result of the paper is the nontrivial evolution of the spectral
function from
a narrow band at $x=0$, to a quasi-noninteracting band at $x
\geq 0.5$. It was also observed 
that the low-energy peaks 
of a cluster spectra acquire finite line-widths as their
energies move away from the chemical potential.

\end{abstract}
\pacs{PACS numbers: 74.20.-z, 74.20.Mn, 75.25.Dw}
\vskip2pc]
\narrowtext

Copper-oxide ladder compounds are currently 
under much investigation~\cite{levy}. Among their
interesting properties are  a spin-liquid ground state
in the undoped limit, and the existence of superconductivity 
upon hole doping~\cite{uehara,science}.
Recently, 
the first angle-resolved photoemission (ARPES) studies of ladder materials
have been reported.
Both the doped and  undoped 
ladder ${\rm Sr_{14} Cu_{24}O_{41}}$ have been 
analyzed,
finding  one-dimensional metallic 
characteristics~\cite{takahashi}. 
Studies  of the ladder compound ${\rm La_{1-x} Sr_x Cu
O_{2.5}}$ found similarities with ${\rm La_{2-x} Sr_x Cu O_4}$, 
including a Fermi
edge~\cite{mizo}. Core-level 
photoemission experiments for ${\rm (La,Sr,Ca)_{14} Cu_{24}O_{41}}$ 
documented its chemical shift
against hole concentration~\cite{core}.
Note that 
the importance  of ARPES studies for other materials such as
 the high-Tc cuprates
is by now clearly established~\cite{shen}. Using this technique
the evolution with doping
of the Fermi surface has been discussed~\cite{marshall}, including the
existence of flat bands near momenta $(0,\pi)-(\pi,0)$~\cite{flat}.

These plethora of experimental results for the cuprates
 should be compared against theoretical predictions. However, the
calculation of the ARPES response even for  simple models 
is a formidable task. The most reliable computational tools for these
calculations are the Exact Diagonalization (ED) method, restricted to
small clusters, and the Quantum Monte Carlo (QMC) technique supplemented by
Maximum Entropy, limited in doped systems
to  high
temperatures due to the sign problem.
Currently, on ladders dynamical properties can be exactly calculated 
at all densities only 
on clusters of size $2 \times
8$~\cite{tsunetsugu,haas,didier2}, 
while the  QMC technique in
the realistic regime of large $U/t$ (Hubbard model) has been applied
on $2 \times 16$ lattices only at half-filling~\cite{endres} and with 1
hole~\cite{endres2}, the latter using an anisotropic ladder since for
the isotropic case the sign-problem is severe.

Due to the limitations of these techniques an
important issue 
still unclear is the
evolution of the one-particle spectral function between the undoped limit,
dominated by antiferromagnetic (AF) fluctuations both
on ladders and planes, and the high hole-density regime where 
those fluctuations are negligible.
While both extreme cases are properly treated by previously available
 numerical methods, the transition 
from one to the other as the hole density $x$ grows is still unknown. This
evolution is expected to be highly nontrivial. For instance,
the presence of hole-pairs in
lightly doped ladders suggests the opening 
of a gap in ARPES, similar to the pseudogap of underdoped
high-Tc cuprates~\cite{dwave}. Shadow-band features in undoped
ladders~\cite{haas}, which are absent at higher hole densities,
 adds to the complexity of this evolution.

Motivated by this challenging problem,
in this paper the density evolution of the spectral function $A({\bf
q},\omega)$ of doped 2-leg $t-J$ ladders
is presented. The calculation
is carried out at zero temperature
on clusters with up to $2 \times 20$ sites, increasing
by a substantial factor the current resolution of the ED techniques.
These intermediate size clusters were reached
 by working with  a small fraction of the total Hilbert space
of the system~\cite{oldtrunca}. The method is variational, although 
accurate as shown below. 
The improvement over previous efforts 
lies in the procedure used to select the basis states of the
problem~\cite{before}. The generation of the new basis is in 
the same spirit as any technique of the
renormalization-group (RG) family. 
If the standard $S_z$-basis is used (3 states per site), experience
shows that
a large number of states is needed to
reproduce qualitatively the spin-liquid characteristics of the undoped
ladders. The reason
is that in 
the $S_z$-basis
 one of the states with the highest weight in the ground state  is still the
N\'eel state, in spite of the existence of a short AF
correlation length $\xi_{AF}$. 
A small basis built up around the N\'eel
state incorrectly favors long-range spin order. However, if the
Hamiltonian of the problem is exactly rewritten in, e.g., the $rung$-basis 
(9 states/rung for the $t-J$ model)
before the expansion of the
Hilbert space is performed, then the tendency to favor
a small $\xi_{AF}$ is natural since one of the dominant states in this basis
for the undoped case
corresponds to the direct product of singlets in each rung,
$|S\rangle$, which has
$\xi_{AF} = 0$ along the chains. Fluctuations of the
Resonant-Valence-Bond (RVB) variety around $|S\rangle$
appear naturally in this new representation of the Hamiltonian leading
to a finite $\xi_{AF}$. Note
that $|S\rangle$ is just $one$ state of the rung-basis, while in the
$S_z$-basis it is represented by $2^{N_r}$ states with $N_r$ the number
of rungs of the 2-leg ladder. In general a few states in the rung-basis
are equivalent to a large number of states in the $S_z$-basis. 
Expanding the Hilbert space
in the new representation is equivalent to 
working in the $S_z$-basis
with a number of states  larger than can be reached
directly with present day computers.
Here 
for simplicity this technique will be referred
to as the Optimized Reduced Basis Approximation (ORBA)~\cite{before}.

\begin{figure}[htbp]
\vspace{-0.5cm}
\centerline{\psfig{figure=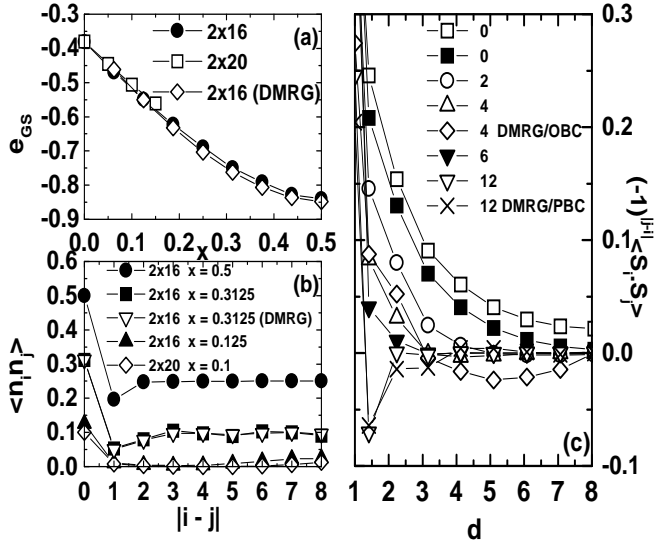,width=9.cm,height=14.0cm}}
\vspace{-6.0cm}
\caption{$e_{GS}$ vs $x$ for the $t-J$ model
using the method described in this
paper for the two clusters indicated $(t=1)$. DMRG results
with OBC are also provided for comparison; (b) Ground state hole-hole density
correlations vs distance for a variety of clusters and densities as
indicated. Some results using DMRG with PBC are also shown;
(c) Ground state staggered spin-spin correlations vs distance $d$
$(= {\bf |i -j|} )$ along
the leg opposite to where site ${\bf i}$ is located. Number
of holes are indicated. Some results with DMRG are also shown. Open
(full) symbols are for $2 \times 16$ ($2 \times 20$) clusters.}
\label{fig1}
\end{figure}

As a first step, let us compare
ORBA predictions for equal-time observables 
against DMRG results\cite{white} for the same clusters.
Here a coupling $J/t=0.4$ is used. Its particular value
is important: if $J/t$ is smaller, then pairs are lost while if it
is larger superconducting correlations are important. Only in a small window
of $J/t$ is that the ground state can be considered as formed by weakly
interacting hole pairs, a regime that we want to investigate in this
paper for its possible connection with the phenomenology of high-Tc at
finite temperature.
Fig.1a contains the ground state energy per site $e_{GS}$ vs $x$ using
$\sim 2-3 \times 10^6$ states in the rung-basis. The 
DMRG energies are obtained with $m=200$ states and open boundary conditions
(OBC). Both sets of data are in good agreement~\cite{detail}.
On $2 \times 20$ clusters, the rung-basis approach allowed us
to study up to 6 holes~\cite{comm21} which has a
full space of $\sim 10^{14}$ states (for zero momentum and total spin),
while the largest previously reported exact study on a $2 \times 10$
cluster and 2 holes needs a  $\sim 5 \times 
10^{5}$ basis~\cite{tsunetsugu,haas}.
Calculating the binding energy, or the chemical
potential $\mu$ vs $x$, from $e_{GS}$ supplemented by the
energies for an odd number of electrons, a  tendency to pair
formation at low hole-density was observed~\cite{tsunetsugu,didier4}.
Fig.1b contains the hole-hole correlations at
several densities,  compared (in one case) with PBC DMRG results.
In Fig.1c  spin-spin
correlations are shown. The rung-basis
properly reproduces the existence of a small $\xi_{AF}$ in the ground
state, that decreases as $x$ grows. 
Size effects are not large for the clusters
studied here, and  good agreement with DMRG 
is observed. 
This technique captures the essence of the
ground state behavior~\cite{sz}. 

\begin{figure}[htbp]
\vspace{-0.5cm}
\centerline{\psfig{figure=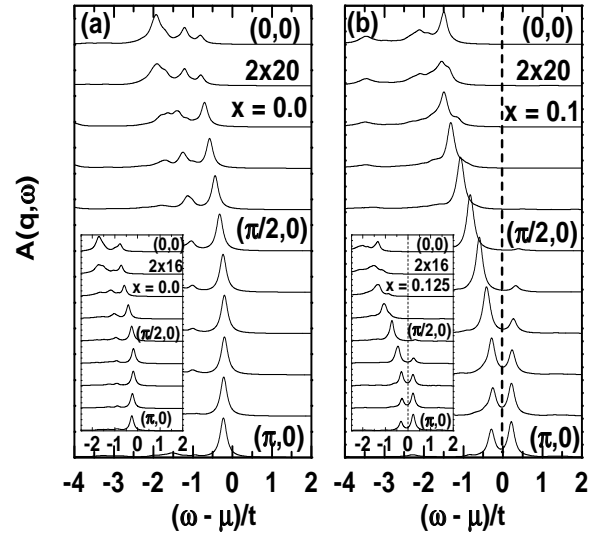,width=9.cm,height=14.0cm}}
\vspace{-6.0cm}
\caption{$A({\bf q},\omega)$ for the bonding band $q_y = 0$
on a $2 \times 20$ cluster. (a) corresponds to $x=0$ and (b) to
$x=0.10$. 
The insets correspond to
$2 \times 16$ clusters.}
\label{fig2}
\end{figure}

To produce dynamical results Ref.\cite{before} was followed, namely
$\sim 10-20\%$ states of the reduced basis $N$-holes ground state
 $| \psi_0 \rangle$ were 
considered~\cite{comm19} and the reduced
subspace with, e.g., $N+1$ holes and
${\bf q}$-momentum was obtained through
 ${\hat O^\dagger_{\bf q}} | \psi_0 \rangle$
(${\hat O^\dagger_{\bf q}} = \sum_{\bf j} e^{i
{{\bf q}\cdot{\bf j}}} {\bar c}^\dagger_{\bf j}$, 
with ${\bar c}^\dagger_{\bf j}$ the hole creation operator 
at ${\bf j}$, dropping the spin-index, in
the rung-basis). 
All states
generated by this procedure were kept, and we worked in such a
 subspace in the subsequent iterations of the continued
fraction expansion~\cite{review}
Only the bonding band subspace
is discussed here~\cite{comm77}. 
The $\delta$-functions have a width $0.1t$ throughout the paper.

Fig.2a corresponds to
the undoped limit. 
A sharp peak is observed at the top of
the PES spectra, maximized at momenta $q_x=7\pi/10$, i.e.
close to the Fermi momentum for noninteracting electrons $q^F_x = 0.66
\pi$. The band defined by those peaks
has a small  bandwidth, as in 2D
models, due to the interaction of the injected holes with the spin
background~\cite{review,quasi}. 
Note
that all peaks at momenta $q_x \geq \pi/2$  carry a similar weight and
the dispersion is almost negligible. This unusual result
is caused by strong correlation effects.
The PES weight above $q^F_x$ , e.g. at $q_x=\pi$,  is
induced by the finite but robust $\xi_{AF}$, and its existence resembles
the antiferromagnetically induced ``shadow'' features discussed before in  
2D models~\cite{shadow}.

Fig.2b contains results at low but finite
 hole-density. Several interesting
details are observed: (i) the PES band near $\mu$
is flat. This should be an 
ARPES observable
result resembling experiments in 2D
cuprates, and it adds to the growing evidence linking the physics of
ladders and planes; (ii) $q_x=\pi$ ($\pi/2$) PES has lost (gained)
 weight compared with $x=0$;
(iii) the total PES bandwidth has increased; and (iv) the IPES
band is intense near $q_x=\pi$, 
and it is separated from the PES band
 by a gap.
The observed gap is $\Delta \sim 0.4t$ and it is caused by hole pairing.
The DMRG/PBC binding energy calculated
for the same cluster and density
is $\sim 0.32t$ ($m=200$, truncation error $\sim 10^{-4}$). 
In the overall energy scale of the ARPES
spectra, this difference is small and does not affect the 
study of the evolution of the dispersion shown here.
Note that the results of Fig.2b are similar to those observed 
near $(\pi,0)$ using ARPES in
the 2D cuprates normal state~\cite{flat,dwave}.

\begin{figure}[htbp]
\vspace{-0.5cm}
\centerline{\psfig{figure=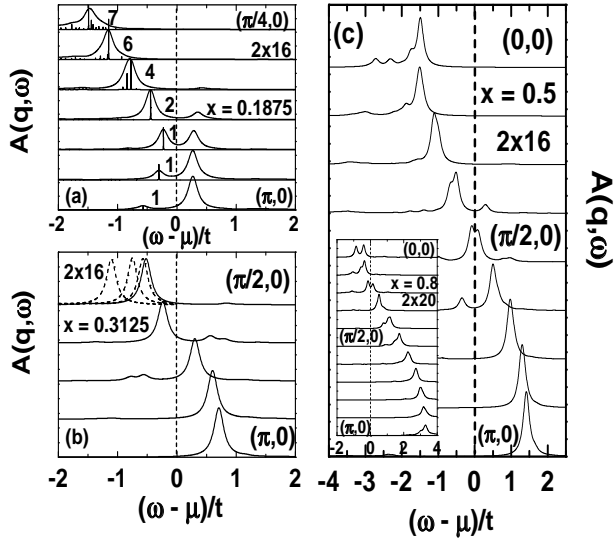,width=9.cm,height=14.0cm}}
\vspace{-6.0cm}
\caption{Same as Fig.2 but using a $2 \times 16$ cluster and the hole
densities and momenta indicated. In (a) PES, the results are also
shown using a width $0.001t$  (and a different
vertical scale) to visualize the individual $\delta$-functions.
The number next to each broad peak
is the number of poles contributing to it (some are difficult
to resolve due to their small weight). In (b) and ${\bf q} = (\pi/2,0)$
results obtained using 1.1, 2.8, 4.5 million states are shown (dashed
lines) from left to right, 
to illustrate the convergence. 
The solid lines were obtained with 6.0 million states.
At this density the
Hilbert space is maximized for the $2 \times 16$ cluster.}
\label{fig3}
\end{figure}

Fig.3a contains results at $x=0.1875$. The trends observed at
$x=0.1$ continue, the more dramatic being the reduction of the
$q_x=\pi$ PES weight caused by the decrease in $\xi_{AF}$.
The lost weight appears in
the  $q_x = \pi$ IPES signal. The gap is still
observed in the spectrum. 
Weak BCS-like features both in PES and IPES near $q^F_x$ can be seen. 
Fig.3b contains data at $x = 0.3125$, and up to $6 \times 10^6$ states.
The Hilbert space is maximized at this density for the $2 \times 16$ cluster.
Now the result resembles more a noninteracting system on a 
discrete lattice.
The IPES signal  is no longer very flat, and
the IPES band now has a clear energy minimum near the momentum
where PES is maximized. 
Fig.3c contains results for $x=0.5$ where
a quasi-non-interacting dispersion is obtained
using about $3 \times 10^6$ states in $|\psi_0 \rangle$. The inset shows
that the trend continues at lower electronic densities. 
The bandwidth
evolves from being dominated by $J$ near half-filling, to having
$t$ as natural scale at $x \sim 0.3$ or larger. 
This evolution is smooth, yet nontrivial, following
the reduction of $\xi_{AF}$ with doping.

A conceptually interesting issue in the context of finite-cluster spectra of
electronic models is whether finite line-widths 
for the dominant peaks can be
obtained by such a procedure. Studying the small clusters reached by 
ED techniques it naively seems that those peaks are usually generated
by just one $\delta$-function (one pole). However, in the bulk limit,
peaks away from the Fermi level should have an intrinsic width. How can
we reach such a limit from finite clusters? 
One possibility is that 
as the cluster grows, the number of poles
 $N_p$ in a small energy window centered at the 
expected peak position must grow also, with their individual intensities
becoming smaller such that the combined strength remains approximately
constant. While this idea seems reasonable, it still has no
explicit verification, but the intermediate size clusters
reached in this study
allow us to test it. Consider as an example
Fig.3a where the actual energy and intensity
 of the poles contributing
to the main features are shown explicitly. As the 
peaks move away from the top of the PES band, $N_p$ was indeed
found to increase
providing evidence compatible with the conjecture made above~\cite{norman}.

Fig.4a contains the main-peak weights in the PES band 
vs density. Size effects are small. The weight at $q_x =
\pi$ diminishes rapidly with $x$, following the
strength of the spin correlations of Fig.1c. 
Overall the region affected the most by
spin correlations is approximately  $x \leq 0.25$.
Fig.4b summarizes the main result of the paper, providing
to the reader the evolution with $x$
of the ladder dominant peaks in $A({\bf q},
\omega)$. 
The area of the circles are proportional to the  peak
intensities. At small $x$ a hole-pairing-induced
 gap centered at $\mu$ is present in the spectrum, both
the PES and IPES spectra 
are flat near $(\pi,0)$, and the band is narrow.
The PES flat regions at high momenta exist also in the undoped limit,
where they are caused by the short-range spin correlations.
Actually 
the resolution in densities and momenta achieved in this study allow us
to reach the conclusion that the undoped and lightly doped regimes 
are smoothly connected.
As $x$ grows to $\sim 0.3$, the flat regions rapidly
loose intensity near $(\pi,0)$, and the gap collapses. 

\begin{figure}[htbp]
\vspace{-0.5cm}
\centerline{\psfig{figure=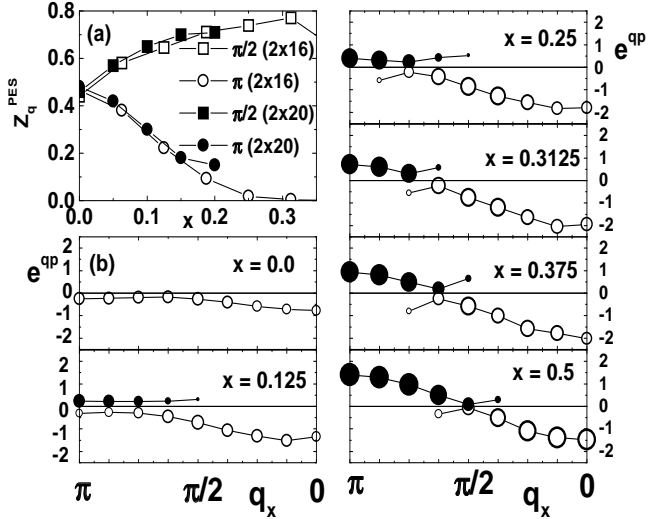,width=9.cm,height=14.0cm}}
\vspace{-6.0cm}
\caption{(a) Weight of the low-energy dominant feature in $A({\bf q},\omega)$
 for two momenta $q_x$ in the bonding band vs $x$.
Cluster sizes are indicated;
(b) Evolution with doping of the dominant feature band ($t=1$).
The open (full) circles are centered at the peak energies 
${\bf e}^{qp}$  below (above) $\mu$.
The hole densities are indicated. The area of the dots is proportional to the
weight of the peak. The results can apply to 2D systems along the
$(\pi,0)-(0,0)$ line.}
\label{fig4}
\end{figure}

The many similarities
between ladders and planes discussed in previous literature~\cite{science} 
suggest that our results may
also be of relevance for 2D systems along the line $(0,0)-(\pi,0)$.
For instance, the abnormally flat
regions near $(\pi,0)$ (Fig.2b)  are similar
to  ARPES
experiments data 
for the 2D cuprates~\cite{flat}, and they should appear in high
resolution photoemission experiments for ladders as well.
Note that in the
regime studied with pairs in the ground state, the flat bands do
not cross $\mu$ with doping but they simply melt.
When $x$ is between 0.3
and 0.4, a quasi-free dispersion is recovered. 
The results of
Fig.4b
resemble a Fermi level crossing at $x \sim 0.3$ and beyond,
while at small hole density no crossing is observed. It is remarkable that
these same qualitative behavior appeared in 
the ARPES results observed recently in  underdoped and overdoped
LSCO~\cite{fujimori}. 
These common trends on ladders and planes suggest
that the large energy scale (LES)
pseudogap ($\sim 0.2 eV$) of the latter~\cite{shen2} 
may be caused by similar
long-lived $d$-wave-like tight hole pairs in the
normal state as it happens in doped ladders, where 
the pairing is caused
by the spin-liquid RVB character of the
ground state~\cite{science}. In the 2D case a similar effect
may originate in the finite $\xi_{AF}$ observed in the underdoped
finite temperature regime (although there is no clear evidence in
the 2D layered materials of a spin-liquid ground state).
A consequence of this idea is that the pairs and LES ARPES
 pseudogap are correlated and they
 should exist as long as $\xi_{AF}$ is non-negligible~\cite{preformed},
a prediction that can be tested experimentally.

Summarizing,  the bonding-band
spectral function of the 2-leg $t-J$ model
 has been calculated, and  
results can be used to guide future ARPES experiments for ladder compounds.
These experiments should observe flat bands and
 gap features near $(\pi,0)$ in the normal
state. The data was found to be
remarkably similar to experimental results for the 
2D cuprates along the $(0,0)-(\pi,0)$ line. 
A common explanation for these features was proposed.
Finally, note that the 
ORBA method discussed here introduces 
a new way to calculate dynamical properties of
spin and hole models on intermediate size clusters.
The method can be applied to a variety of strongly correlated electronic
models.

The authors specially thank J. Riera for many useful suggestions.
The financial support of the NSF grant DMR-9520776,  
CNPq-Brazil, CONICET-Argentina, and the NHMFL
In-House Research Program (DMR-9527035) is acknowledged.

\medskip

\vfil


\begin{references}
\bibitem{detail} For the $2
\times 16$ cluster, the Hilbert space is maximized for 10 holes reaching
$\sim 10^{12}$ states. In this
situation $e_{GS}$ was
0.76 with ORBA, and 0.768 with the DMRG method ($m = 200$) 
using this time
 periodic boundary conditions (PBC) in both cases. The error is in the
last digit.
\bibitem{comm21} For densities $ 0.2 \leq x \leq
0.5$, where the size of the Hilbert space is the largest,
 the convergence on the $2 \times 20$ cluster
was not good enough for our purposes.
\bibitem{sz} This does not occurs in the $S_z$-basis which
favors long-range order both in the spin and charge channel.
\bibitem{comm19} In practice, we observed that the shape of the
spectral functions
are very similar irrespective of
whether a small fraction of $|\psi_0 \rangle$ is used
or the full state. They differ only in an energy shift. Thus, most of the
$qualitative$ features of the ARPES spectra can be obtained with only
a fraction of the states used in the present paper.
\bibitem{comm77} Antibonding band results will be shown elsewhere.
\bibitem{quasi} Since the ARPES peaks in the 2D cuprates
 for some momenta and densities
are so broad that they cannot be labeled as
quasiparticles, we will here avoid using 
that language also for the case of ladders.
\end{references}
\end{document}